\documentclass[aps,epsfig,preprint,groupedaddress]{revtex4-1}
\usepackage{amssymb}
\usepackage{graphicx}
\usepackage{color}
\usepackage{hyperref}
\usepackage{fullpage}
\usepackage{comment}
\makeatletter
\gdef\@ptsize{0}
\let\@currsize\normalsize
\makeatother
\usepackage{setspace}
\doublespace

\begin{document}
\title {Radiation Reaction Effect on Laser Driven Auto-Resonant Particle Acceleration}
\author{Vikram Sagar, Sudip Sengupta and P.K.Kaw}
\affiliation{Institute for Plasma Research, Bhat, Gandhinagar - 382428, India}
\begin{abstract}
The effects of radiation reaction force on laser driven auto-resonant particle acceleration scheme are studied using Landau-Lifshitz equation of motion. These studies are carried out for both linear as well as circularly polarized laser fields in the presence of static axial magnetic field. From the parametric study, a radiation reaction dominated region has been identified in which the particle dynamics is greatly effected by this force. In the radiation reaction dominated region the two significant effects on particle dynamics are seen viz., (1) saturation in energy gain by the initially resonant particle, (2) net energy gain by a initially non-resonant particle which is caused due to resonance broadening. It has been further shown that with the optimum choice of parameters this scheme can be efficiently used to produce electrons with energies in the range of hundreds of TeV. The quantum corrections to the Landu-Lifshitz equation of motion have also been taken into account. The difference in the energy gain estimates of the particle by the quantum corrected and classical Landu-Lifshitz equation are found to be insignificant for the present day as well as upcoming laser facilities.   
\end{abstract}

\maketitle
\section{Introduction}
The rejuvenated interest in the Auto-Resonant scheme which in principle can cause large unbound acceleration of particle is due to the concurrent developments in the ultra high magnetic field generation experiments and laser technology. The advancement in these areas have lead to the availability of non-destructive magnetic fields of the order of ($\sim 100$ Tesla) \cite{100-T,100-Ta,magnetic} together with lasers having intensities of the order of $10^{22} W/cm^2$ \cite{Laser-1}  which are expected to increase even further \cite{Laser-2,Laser-3,Laser-4}.  The scheme of Auto-Resonant particle acceleration was discovered by Kolomenski{\"I} and Lebedev \cite{ref-1,ref-1a} and, independently, by Davydovski{\"I} \cite{ref-2}. The previous studies[\cite{ref-3,ref-4,ref-5,ref-6,ref-7,ref-8,ref-9,ref-10,ref-11,ref-12,ref-13,ref-14,ref-15,ref-16,ref-17,ref-18,ref-19}] on the topic have not considered the effect of radiation reaction in this scheme. The radiation reaction (or radiative friction) is a recoil force acting on the accelerating (retarding) particle as a result of emission of radiation and can in turn influence the motion of the particle. In the absence of radiation reaction it is known that there is no net transfer of energy to the particle interacting with the laser and non-resonant magnetic field. However it has been recently shown from exact analytical results \cite{recent-1} that the inclusion of radiation reaction which has an amplitude dependent coefficient of friction results in the net transfer of energy to the particle interacting with laser field. In this context, the theoretical understanding of the effect of radiation reaction force on the Auto-resonant scheme can be of central importance in the optimum use of these laboratory generated ultra high magnetic fields and lasers for the experimental realization of the scheme. 

 In the Auto-Resonant acceleration scheme, the particle is subjected to the combined field of laser and static axial magnetic field. The particles are accelerated as a result of self-sustained resonance between the particle and the laser field due to which the particle remains phase locked with the laser field. The resonant phase locking condition can be expressed as, $\Gamma-P_{x}-\Omega_{c}=0$, where $\Omega_{c}(=\frac{eB_{o}}{mc\omega})$ is the cyclotron frequency normalized to laser frequency $\omega$, $\lq P_{x}/mc\rq$ is the normalized longitudinal component of particle four momentum and $\lq \Gamma \rq$ is the relativistic factor. The mechanism for particle acceleration by auto-resonance scheme can be  briefly described in a following way: the particle initially at rest and satisfying the initial resonance condition ($\omega=\Omega_{c}$) is accelerated along the electric field of laser. The increase in particle energy along the electric field component in the transverse direction results in relativistic mass increase which in turn lowers the cyclotron frequency of the particle. The transverse velocity gain results in the action by the magnetic field component of the wave, which pushes the particle along the direction of the propagation of the wave. The relativistic velocity acquired by the particle along the longitudinal direction results in a Doppler shift to a lower frequency of the wave as {\lq\lq seen\rq\rq} by the particle. In this case, the Doppler shifted wave frequency to the lower frequency equals to the relativistically modified  cyclotron frequency, and the particle remains {\lq\lq synchronously\rq\rq} in cyclotron-resonance condition. This result in continuous increase in particle energy and momentum along the direction of propagation of the wave.

  The present work takes into account the effect of radiation reaction force (or radiative friction) on the dynamics of a particle in the combined field of intense laser and static axial magnetic field using the Landau-Lifshitz (LL) equation\cite{LL}. The choice of LL equation over the Lorentz-Abraham-Dirac (LAD) equation in the study of particle dynamics is because of the reason that it circumvents the problems such as violation of principle of causality and existence of nonphysical runway solutions associated with the LAD equation\cite{LAD}.  The LL equation can be perturbatively derived from LAD equation \cite{LL} and thus is equivalent to it in first order. This equation has also been used in the recent laser matter interaction studies \cite{recent-1,recent-2,recent-3,recent-4,recent-5,recent-6,recent-7,recent-7,recent-8,recent-9} to highlight the effect of radiation reaction force on particle dynamics for the present day as well for upcoming laser facilities.  

 In the present work radiative friction dominated parametric regime has been explored to study the dynamics of initially resonant as well as non-resonant particle interacting with the combined field of laser and static axial magnetic field. In this regime, at first the energy gain of the initially resonant particle is estimated at which the radiation friction effects on the particle dynamics become prominent for the present day as well as upcoming laser facilities. For the initially non-resonant particle it is well known that in the absence of radiative friction there is no net transfer of energy at the end of each gyration. In the present study the effect of de-phasing caused by amplitude dependent radiative friction on the initially non-resonant particle from the laser field has also been investigated in the presence static axial magnetic field.  A detailed comparative study has been carried out on the peak energy gain of the particle by linearly and circularly polarized lasers. Further the effect of quantum correction \cite{Quantum-Correction,Quantum-Correction-1} in the LL equation of motion has also been considered for estimating the energy gain by the particle. The optimum conditions have been predicted for the large peak energy gain by the particle.         

The organization of the paper is as follows: In section (II), mathematical frame work for studying the particle dynamics in the combined field of laser and static axial magnetic field is presented. In section (III), the Landau Lifshitz equation of motion is numerically solved using R.K(Runge-Kutta) method for Linear and Circular polarization of the laser. Section (IV), contains the summary and discussion of the results.

\section{Landau-Lifshitz equation for particle motion in the \\ combined field of laser and static axial magnetic field}
The effect of radiation damping on relativistic particle dynamics is described by the Lorentz-Abraham-Dirac (LAD) equation. 
\begin{eqnarray}
mc\frac{du^{i}}{d\tau}&=&\frac{e}{c}F^{ik}u_{k}+g^{i}\\ 
\frac{dx^{i}}{d{\tau}}&=&u^{i}
\end{eqnarray}
Here $u^{i}=(\Gamma,{\bf P}/mc)$ is the four velocity, $F_{ik}=\partial_{i}A_{k}-\partial_{k}A_{i}$ is the EM field tensor with $A_{i}=(\Phi,\vec{A})$ being the four potential. The radiation friction force in the LAD equation \cite{LAD,LL} is given by,
\begin{eqnarray}
g^{i}&=&\frac{2e^{2}}{3c}\big(\frac{d^{2}u^{i}}{d\tau^{2}}-u^{i}u^{k}\frac{d^{2}u_{k}}{d\tau^2} \big)
\end{eqnarray} 
The LL radiation friction force is derived as a perturbation to the equations of motion, this approximation is valid provided there exists a frame of reference, where the LL radiation friction force is small compared to the Lorentz force. Using the equation of motion \cite{LL} (neglecting $g^i$ in the first approximation), $\frac{du^i}{d\tau}$ and $\frac{d^2u^i}{d\tau^2}$ are evaluated as
\begin{eqnarray}
\frac{du^i}{d\tau}&=&\frac{e}{mc^{2}}F^{ik}u_{k}\\
\frac{d^{2}u^{i}}{d\tau^{2}}&=&\frac{e}{mc^{2}}\frac{\partial F^{ik}}{\partial x^{i}}u_{k}u^{l}+\frac{e^{2}}{m^{2}c^{4}}F^{ik}F_{kl}u^{l}
\end{eqnarray}
Substituting the above expression in the Eq [3] the radiation friction force in the LL form stands as 
\begin{eqnarray}\nonumber
g^{i}&=&\frac{2e^{3}}{3mc^{3}}\frac{\partial F^{ik}}{\partial x^{l}}u_{k}u^{l}-\frac{2e^{4}}{3m^{2}c^{5}}F^{il}F_{kl}u^{k}+\frac{2e^{4}}{3m^2c^5}(F_{kl}u^{l})(F^{km}u_{m})u^{i}
\end{eqnarray}
Using the above expression for $g^i$, the LL equation\cite{LL} of motion is given by
\begin{eqnarray}
mc\frac{du^{i}}{d\tau}&=&\frac{e}{c}F^{ik}u_{k}+\frac{2e^{3}}{3mc^{3}}\frac{\partial F^{ik}}{\partial x^{l}}u_{k}u^{l}-\frac{2e^{4}}{3m^{2}c^{5}}F^{il}F_{kl}u^{k}+\frac{2e^{4}}{3m^2c^5}(F_{kl}u^{l})(F^{km}u_{m})u^{i}
\end{eqnarray}
     
The 3-D spatial part of the radiative friction corresponding to additional terms in the momentum equation in the LL equation is 
\begin{eqnarray}
f&=&\frac{2e^{3}}{3mc^{3}}\big(1-\frac{v^2}{c^2}\big)^{-1/2} \Big\{ \big(\frac{\partial}{\partial t}+\vec{v}.\nabla \big)\vec{E}+\frac{1}{c}\vec{v}\times\big(\frac{\partial}{\partial t}+\vec{v}.\nabla \big)\vec{H} \Big\}\\ \nonumber
&&+\frac{2e^{4}}{3m^{2}c^{4}}\Big\{\vec{E}\times\vec{H}+\frac{1}{c}\vec{H}\times(\vec{H}\times\vec{v})+\frac{1}{c}\vec{E}(\vec{v}.\vec{E})\Big\}\\ \nonumber
&&-\frac{2e^4}{3m^2c^{5}(1-\frac{v^{2}}{c^{2}})}\vec{v}\Big\{(\vec{E}+\frac{1}{c}\vec{v}\times{\vec{H}})^2-\frac{1}{c^2}(\vec{E}.\vec{v})^{2}\Big\}
\end{eqnarray} 
In the following section the numerical calculations are carried out using all the three terms in the radiation reaction force given by above equation. However from the order comparison as well as in the numerical simulations the third term has been found to be the most dominant term.

  In the absence of radiative friction the inclusion of static axial magnetic field in Eq.[1] couples the particle dynamics along the transverse spatial components and decoupling of the motion yields driven harmonic oscillator equations along these components\cite{ref-17,ref-19}. The oscillators are driven by externally applied field of laser together with the static axial magnetic fields. The inclusion of radiative friction causes the dampening of these oscillators and non-linear coefficient of friction depends upon the amplitude of the applied field.    

In the present work, the quantum electrodynamics weakening of the radiation friction \cite{Quantum-Correction,Quantum-Correction-1} has also been taken into account by multiplying the radiative friction part of LL equation with a form-factor.The form-Factor is given by $G_{e}({\chi_{e}})$, where $\chi_{e}(=\frac{\sqrt{(F_{\mu\nu P_{\nu}})^2}}{E_{s}mc})$ is the relativistic and gauge invariant parameter characterizing the probability of the gamma-photon emission by the electron with four momentum $P_{\nu}$ and $E_{s}(=\frac{m_{e}^3c^{3}}{e\hslash})$ is the $\lq  \text{Sauter-Schwinger}\rq$  field. The 3D spatial part the parameter $\chi_{e}$ is given by 
\begin{equation}
\chi_{e} = \frac{\Gamma}{E_{s}}\sqrt{((\vec{E}+\frac{1}{c}\vec{v}\times{\vec{H}})^2-\frac{1}{c^2}(\vec{E}.\vec{v})^{2})}
\end{equation} 
For numerical calculations \cite{Quantum-Correction-1} the corresponding form-factor is given by 
\begin{equation}
G(\chi_{e})\approx\frac{1}{(1+8.93 \chi_{e}+2.41 \chi_{e}^2)^{2/3}}
\end{equation}

The above equations can be simplified by introducing a constant $\tau_{0}$, with dimension of time and is given by $\tau_{0}=\frac{2}{3}\frac{e^2}{mc^3}$. For electron/positron it has a numerical value $\tau_{0}=6.24 \times 10^{-24} s$. 

The equation of motion is numerically integrated by expressing the variables in dimensionless form by using the following normalizations:
$\vec{R}(x,y,z)\rightarrow k\vec{R}, t\rightarrow\omega t, \vec{P}\rightarrow\frac{\vec{P}}{mc}, \Gamma\rightarrow\frac{\Gamma}{mc^{2}}, B\rightarrow\frac{qB}{m\omega c}, E\rightarrow\frac{qE}{mc\omega}, \hat{A}\rightarrow\frac{{q}A}{mc^2},\Omega_{c}\rightarrow \frac{qB_{0}}{mc\omega} $. The vector potential of an elliptically polarized laser traveling along $\hat{x}$ direction in the presence of a constant homogeneous axial magnetic field is given by, 
\begin{equation}
\vec A=\left(A_{0}(1-\kappa^{2})^{1/2}\sigma(\xi)-\frac{B_{0}z}{2}\right)\hat{y}+ \left(A_{0}\kappa \sigma_{1}(\xi)+\frac{B_{0}y}{2}\right)\hat{z}
\end{equation}
in the above expression $\kappa$ defines the polarization state of the laser for linear polarization ($\kappa = 0,\pm 1$) and the circular polarization of the laser is defined by ($\kappa = \pm\frac{1}{\sqrt{2}}$). The other terms as specified above are: the phase of the laser is given by $\xi=(\omega t- k x)$, $\sigma(\xi) =(Sin(\xi))$ and $\sigma_{1}(\xi)(=Cos(\xi))$ are the oscillatory parts, $A_{0}$ is the peak laser amplitude and $B_{0}$ is the magnitude of the external magnetic field. The equation of motion has been integrated in terms of co-ordinate $\xi$.

\section{RESULTS OF NUMERICAL SOLUTION OF THE LL EQUATION OF MOTION} 
The dynamics of charged particle in the combined field of laser and static axial magnetic field is studied taking radiation reaction force into account.  The dynamics of the particle is obtained by numerically solving the LL equation of motion using Runge-Kutta integration scheme. The particle dynamics is studied for both linearly and circularly polarized laser. The numerical results have been validated by comparing them with the exact analytical results \cite{recent-1} for particle dynamics in the linear and circularly polarisied laser in the absence of axial magnetic field. The results obtained by solving the LL equation are compared with that of the known results of Lorentz force equation. In the following study for the sake of convenience the previously expressed resonance condition has been re-arranged and expressed in terms of parameter $\lq r(=\frac{{\Omega}_{c}}{\Delta \omega} )\rq $, which is the ratio of cyclotron frequency $\lq \Omega_{c} \rq$ to the product of laser frequency $\lq \omega \rq$ with a variable $\lq \Delta(=\Gamma-P_{x}) \rq$. The resonant interaction defined by $\lq r \rq =1$.   

 The configuration and momentum space trajectories of the initially resonant particle in the absence as well as presence of the radiative friction in the field of circularly polarized laser are shown in Fig-1. From the figure it can be seen, that the radiative friction is not prominent in the initial part of interaction. The resonant interaction accelerates the particle along the direction of propagation of laser which results in the increase in its orbit size for both momentum and configuration space trajectories. On comparing the momentum space particle trajectory in sub-plot (A) and (B), it is found that the effect of radiation reaction force on the particle dynamics becomes pronounced with the increase in particle momentum. The prominence of radiative friction causes a marked difference between the momentum as well as configuration space trajectories which are shown in sub-plots (A) and (C) respectively. From the figure it can be inferred that the presence of radiative friction in turn pushes the particle out of resonance which results in the decrease of the particle momentum. The decrease in particle momentum results in the shrinking of its corresponding configuration space trajectory in the transverse plane.   

The comparison of the effect of inclusion of radiative friction on the initially non-resonant particle trajectories (i.e $r=.7$) in the momentum and configuration space to that of its absence is shown in Fig-2. From sub-plots A and B, it is evident that in contrast to the case when the radiation friction in neglected and there is no net transfer of momentum to the particle, the inclusion of the radiative friction results in the gradual de-phasing of the particle from the applied laser and magnetic field. The de-phasing of the particle results in net gradual increase in particle momentum which in turn pushes the particle to the cyclotron resonance and resulting in large increase in longitudinal particle momentum. The change (increase and decrease) in particle momentum results in the corresponding variation in the size of particle orbit in both momentum and configuration space trajectories. The resonant particle as in the earlier case is then gradually pushed out of the resonance.   

 From the both trajectories it can be observed that the radiative friction causes the de-phasing of the particle from the laser field resulting in gradual increase in the particle momentum and increase in its orbit size. The increase in longitudinal momentum causes a gradual Doppler down shift in frequency of the laser as $\lq \text{seen} \rq$ by the particle and thus driving it in cyclotron resonance with the applied fields. The resonant interaction results in rapid increase in particle momentum. The continuous evolution of resonance parameter beyond resonant condition given by $\lq r \rq =1$ drives the particle out of resonance with applied fields resulting in fall of particle momentum and its eventual saturation.  
 
 In the Fig-3, the plot of the energy gain as a function of null co-ordinate $\lq \xi(=\hat{t}-\hat{x}) \rq$ is shown for the initially resonant as well non-resonant particle with the applied laser and static axial magnetic field. From the sub plots A and B of the figure, it can be inferred that the peak energy gain in the case of initially resonant particle for a given laser intensity is larger in the case of circularly polarized laser than in comparison to the linear polarization. The radiative friction assisted de-phasing of particle from the laser field results in the increase of its energy which is found to be more in the case of circular polarized laser. Thus it can be inferred that for the same initial laser intensity and static axial magnetic field, the circularly polarized laser drives the initially non-resonant particle into cyclotron resonance at rate faster than linearly polarized laser. The acceleration of initially non-resonant particle caused by its de-phasing from the laser field due to radiative friction is of great significance. This significantly relaxes initial condition in terms of the precise requirement of applied axial magnetic field for the resonant acceleration of the particle that has previously hindered the experimental realization of the scheme. 

 The effect of radiation reaction force on the characteristic resonance curve is shown in the Fig-4. The three energy plots as function of the resonant parameter $\lq r \rq$ in the figure corresponds the same normalized static axial magnetic field $(.83)$ and three different normalized values of the laser amplitudes given by $A_{0}=10,300$ and $700$. From the figure it can be observed that the presence of radiation reaction force broadens the width of the resonance curve as well as causes a modification in the resonant condition. The resultant resonant peaks are shown in the Fig-4 and from the figure it can be inferred that the width of the resonance curve as well as the shift in resonant peaks defined in terms of $\lq r \rq$ depends upon the laser amplitude. The shifting in the resonance peaks is similar in its effect to that of a damped driven oscillator with a variable coefficient of friction. In the present case the oscillator is driven by the axially applied magnetic field together with laser and the dampening is caused by radiative frictional coefficient which has non-linear dependence upon laser amplitude.       
  
  One of the major challenge in the experimental realization of the scheme is the requirement of the static magnetic fields of the order of ($\sim 10^{4}$ Tesla) which has been shown to decrease by one order ($\sim 10^3$ Tesla) by inclusion of radiative friction. In Fig-5, it shown that significant reduction in magnetic field which are of the order of presently available magnetic fields ($\sim 10^2 $ Tesla) in the laboratory  and further gain in particle energy can be simultaneously achieved by imparting a small longitudinal momentum ($\sim $ Mev) to the initially non-resonant particle. From the sub-plots A,B,C and D it can be inferred that for the same initial value of resonant parameter $\lq r(=.95) \rq$, an approximate six times gain in energy ($\approx 100$ TeV) can be achieved by four times increase in the initial longitudinal momentum($\approx 2.5-10$MeV). On comparing the gain in energy of the particle for linear and circular polarization it is evident that the energy gain is more for the circular polarization.  

The initial longitudinal momentum imparted to the particle causes a Doppler down shift in the laser frequency as $\lq \text{seen} \rq$ by the particle. As a result the particle remains phase locked in the accelerating phase of the laser resulting in large gain in peak energy. In Fig-6, it shown that for initially non-resonant particle the peak energy gain depends upon the rate of evolution of resonant parameter $\lq r \rq$. On comparing figures (5 and 6), the peak energy gain is shown to be more for the slower evolving resonant parameter. The rate of evolution of resonant parameter $\lq r \rq$ is shown to have inverse dependence upon the initial longitudinal particle momentum. In Fig-7, it is further shown that the peak energy gain increase linearly with the normalized laser amplitude for both linear and circular laser polarization. The quantum correction defined by equation Eq-8 and Eq-9 to the radiation force in the LL equation of motion are found to be negligible for various peak energy gains by the particle shown in figure (5 and 6) and corresponding to different parameters. Further a parametric space has been identified in this acceleration scheme where the quantum corrected energy estimates differ from the one obtained from classical estimates. The effect of quantum corrections are shown in Fig-8, which for circularly polarized laser correspond to very high laser intensities ($\sim 10^{25} W/cm^{2}$).    

 In this study the energy gain by the particle can be optimized in terms of the laser parameters i.e its polarization and amplitude along with the initial value of resonant parameter $\lq r \rq$, which can be altered by varying particle's initial longitudinal momentum and applied static axial magnetic field. One of the possible experimental ways of extracting the energetic particle from laser field is by using a metallic filter which can absorb the laser light and whose thickness is much smaller than stopping distance of the particles\cite{Particle-Stopping}.

\section{Summary}
 The interaction of charged particle with the combined field of ultra intense laser and static axial magnetic field has been explored using Landu-Lifshitz equation. It has been shown that the radiation reaction force (or radiative friction) plays a significant role in the laser driven Auto-resonant scheme of particle acceleration. The interaction of the stationary particle with fields is studied at first, by considering it to be initially in cyclotron resonance with applied fields and secondly, by choosing it to be initially non-resonant. For the initially resonant case, it has been shown that in contrast to unbounded increase in particle energy in the absence of radiation reaction force, the particle energy saturates with the inclusion of radiative friction. For the second case corresponding to initially non-resonant particle, the inclusion of radiation friction results in the net increase of longitudinal particle momentum at each gyration which in turn pushes it to cyclotron resonance resulting in large energy gain ($\sim$ TeV). The radiative friction eventually causes the de-phasing of the particle from the laser field resulting in net saturated increase in the particle energy. It has been shown the presence of radiative friction results in the resonance width broadening as well in the shifting of resonance peak to the higher values of the resonant parameter $\lq r\rq$. The broadening in the resonance width and its shifting is shown to depend upon the laser amplitude. The fact that the radiative friction assists in the acceleration of initially non-resonant particle is of great significance and is expected to play a role in the experimental realization of the scheme. Further, the overall peak energy gain of the particle is shown to have an inverse dependence upon the rate of evolution of previously defined parameter $\lq r \rq$.  

   In terms of parameter optimization, it has been shown that the peak energy gain of the particle for a laser ($\lambda \sim 1 \mu$m ) with same intensity ($\sim 10^{21} W/{cm}^2$) can further be enhanced $(\sim $ 100 TeV) by imparting a small longitudinal momentum $(\sim $ 2.5-10 Mev) to the initially non-resonant particle prior to the onset of interaction. The initial longitudinal momentum causes a Doppler down shift in the laser frequency as $\lq \text{seen} \rq$ by the particle and in turn resulting in up to two order reduction in the applied static axial magnetic field requirements  $(\sim 200-400$ Tesla) for the cyclotron resonance and are available in laboratories. It has also been demonstrated that the peak energy gain by the particle grows linearly with the laser amplitude. The quantum corrections in the energy estimates arising due to the weakening of the radiative friction are found to be negligible for the above set of parameters. It has been shown that the pronounced quantum corrections to the energy estimates in this acceleration scheme will be begin at very high laser intensities ($\geq 10^{25}W/{cm}^2$).      
    

\newpage

\begin{center}
{\bf FIGURE CAPTIONS}
\end{center}

\noindent
Fig(1): The momentum and configuration space particle trajectories for the initially resonant in combined field of Circularly Polarized laser (CP) together with static axial magnetic field with and without radiation reaction force. 

\noindent
Fig(2):The momentum and configuration space particle trajectories for the initially non-resonant in combined field of Circularly Polarized laser (CP) together with static axial magnetic field with and without radiation reaction force. 

\noindent
Fig(3): The energy gain of the initially resonant as well as non-resonant particle in combined static axial magnetic field along with that of linearly (LP) and circularly polarized laser (CP) respectively.

\noindent
Fig(4): Resonant curve for different laser amplititudes.

\noindent
Fig(5): Energy gain of the initially non-resonant particle as function of resonant parameter $\lq r \rq$ for linearly polarizied (LP) and cricularly polarizied laser (CP). 

\noindent
Fig(6): Evolution of resonant parameter $\lq r \rq$  for different initial values of Longitudinal momentum for lineraly (LP) and circularly polarizied (CP) lasers. 

\noindent
Fig(7): Peak energy gain as a function of laser amplitude ($A_{0}$). 

\noindent
Fig(8): The comparison of classical and quantum corrected energy estimates. 

\newpage
\begin{figure}
\begin{center}
\includegraphics[angle=-90, width=1\textwidth]{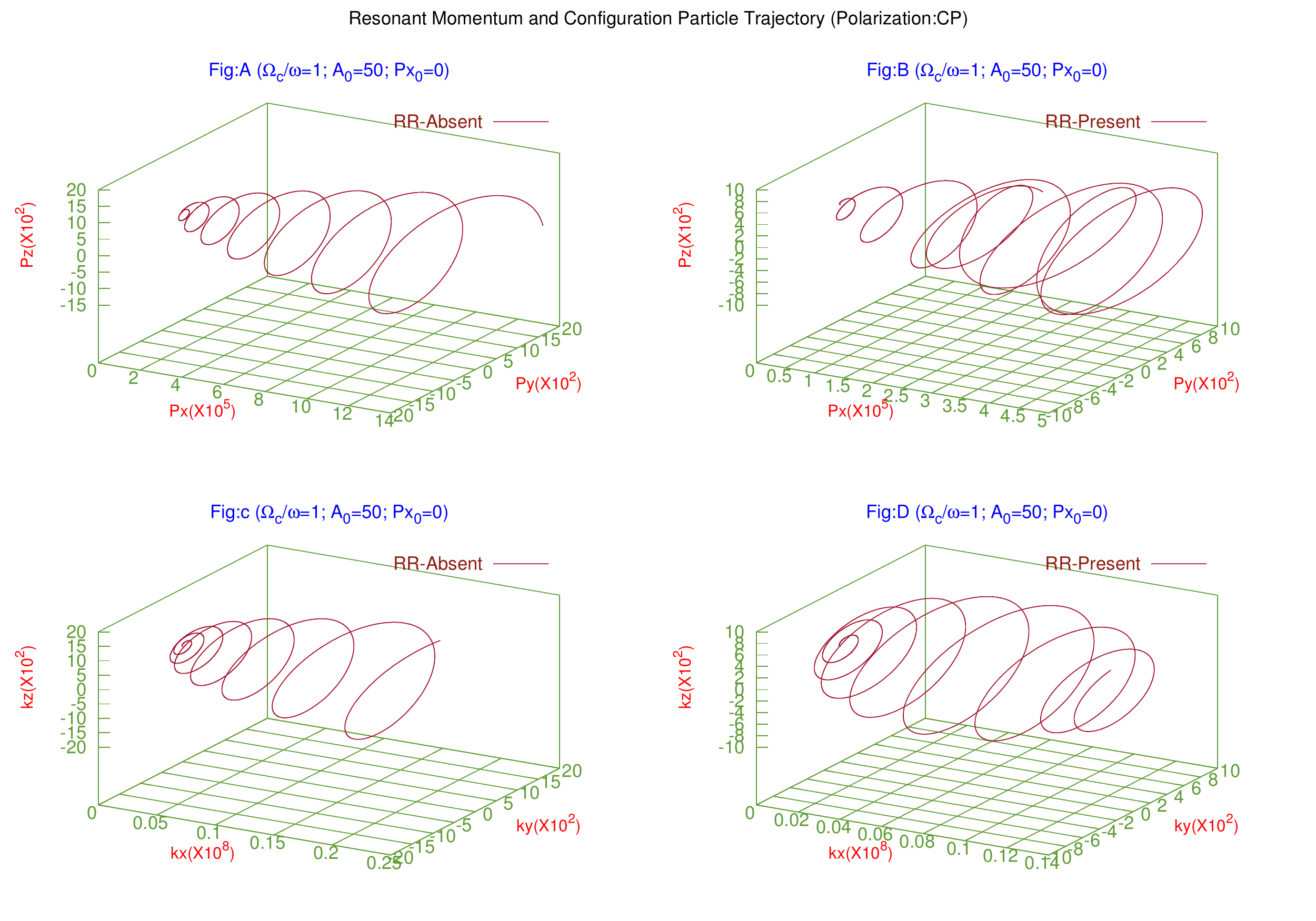}
\caption{The momentum and configuration space particle trajectories for the initially resonant in combined field of Circularly Polarized laser (CP) together with static axial magnetic field with and without radiation reaction force.}.
\end{center}
\end{figure}

\newpage
\begin{figure}
\begin{center}
\includegraphics[angle=-90, width=1\textwidth]{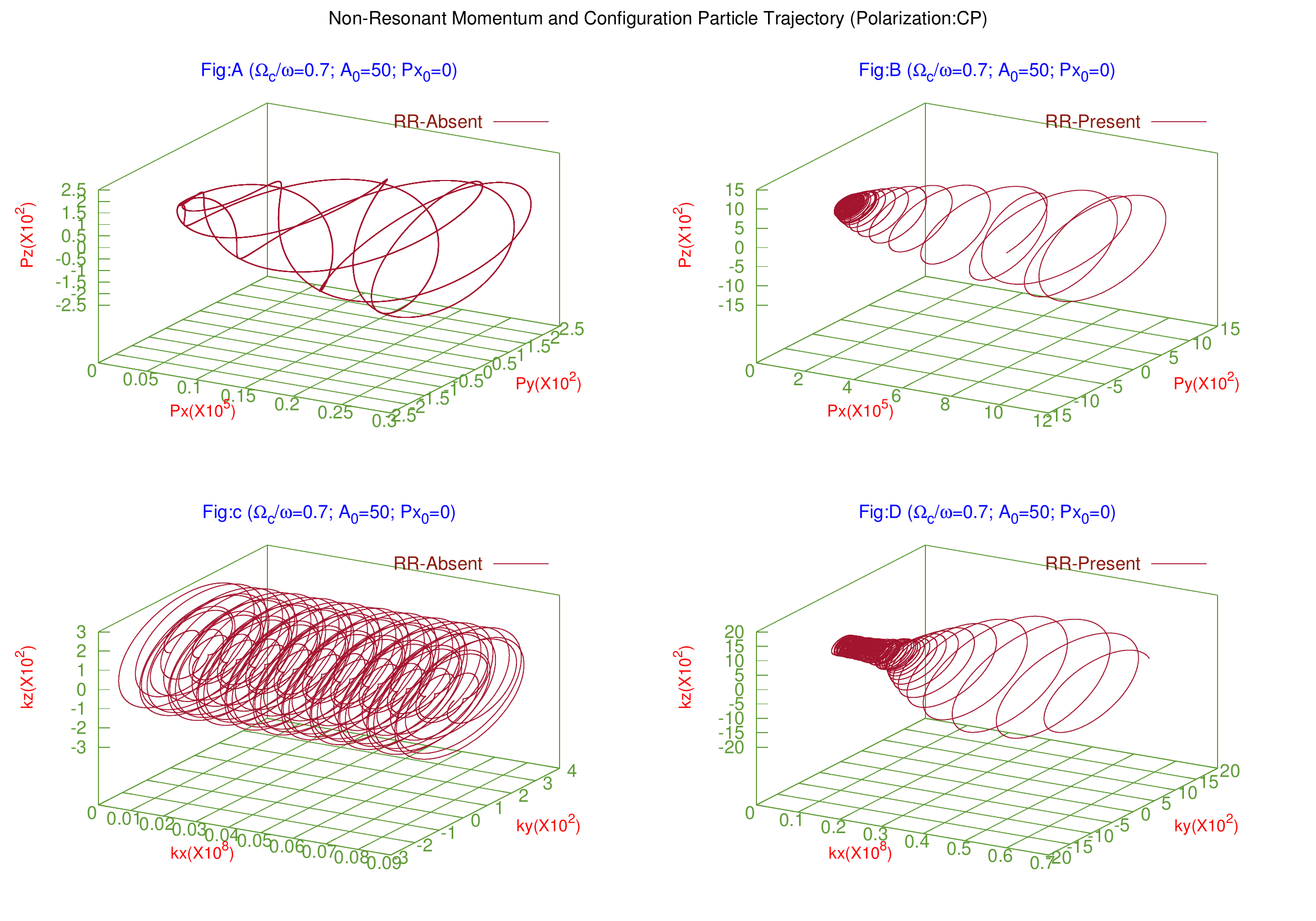}
\caption{The momentum and configuration space particle trajectories for the initially non-resonant in combined field of Circularly Polarized laser (CP) together with static axial magnetic field with and without radiation reaction force.} 
\end{center}
\end{figure}

\newpage
\begin{figure}
\begin{center}
\includegraphics[angle=-90, width=1\textwidth]{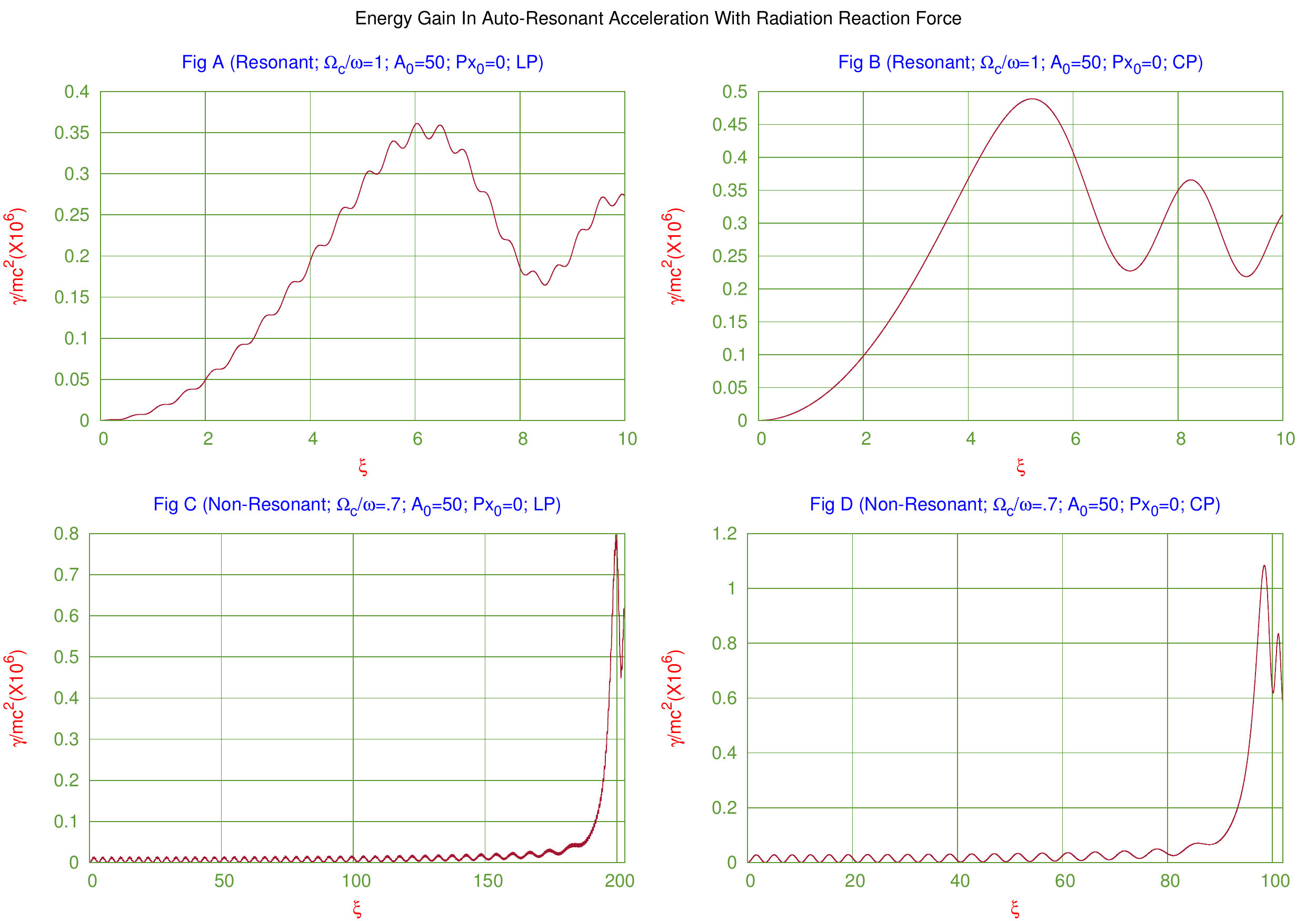}
\caption{The energy gain of the initially resonant as well as non-resonant particle in combined static axial magnetic field along with that of linearly (LP) and circularly polarized laser (CP) respectively.}
\end{center}
\end{figure}

\newpage
\begin{figure}
\begin{center}
\includegraphics[angle=-90, width=1\textwidth]{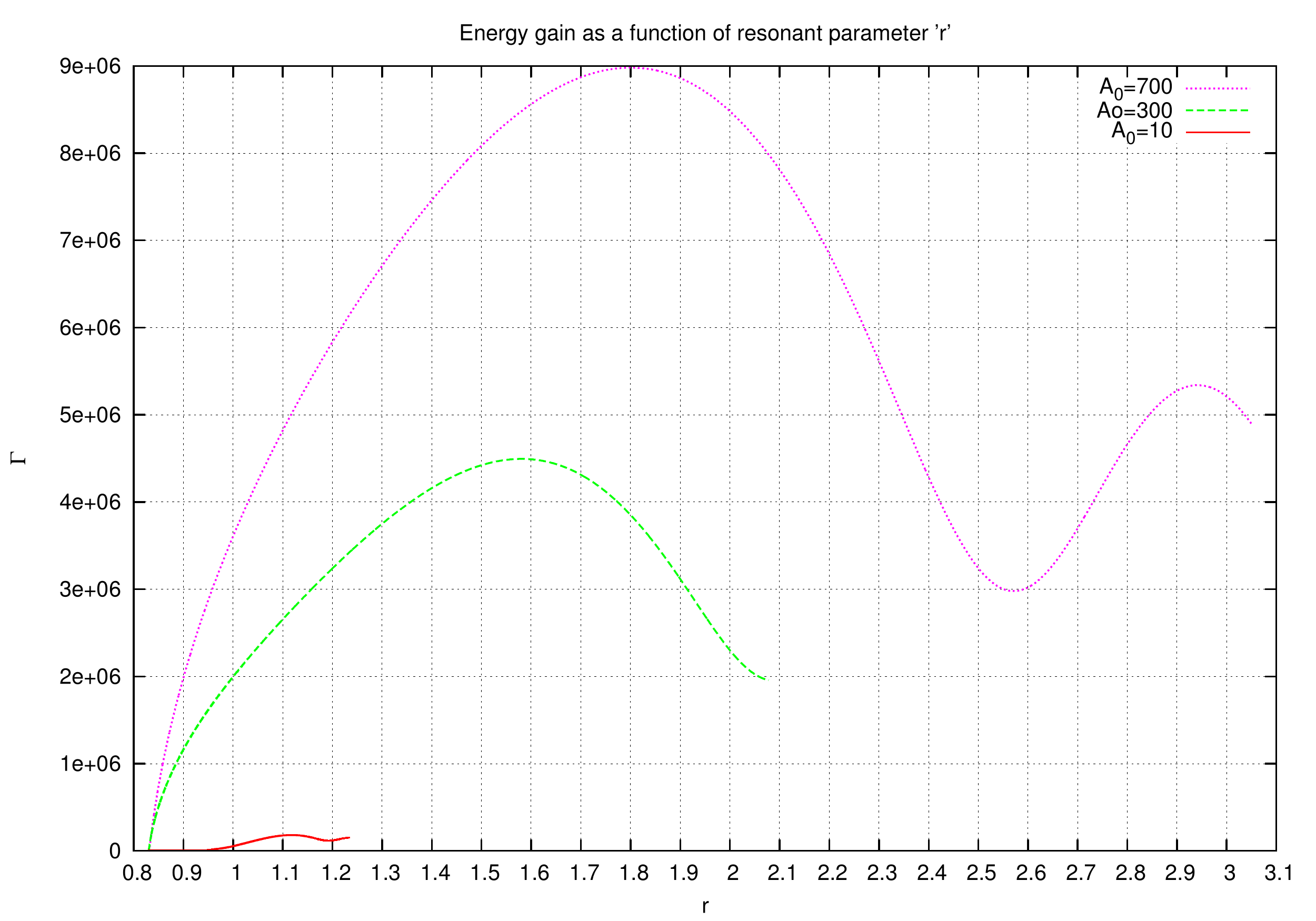}
\caption{Resonant curve for different laser amplititudes.}
\end{center}
\end{figure}

\newpage
\begin{figure}
\begin{center}
\includegraphics[angle=-90, width=1\textwidth]{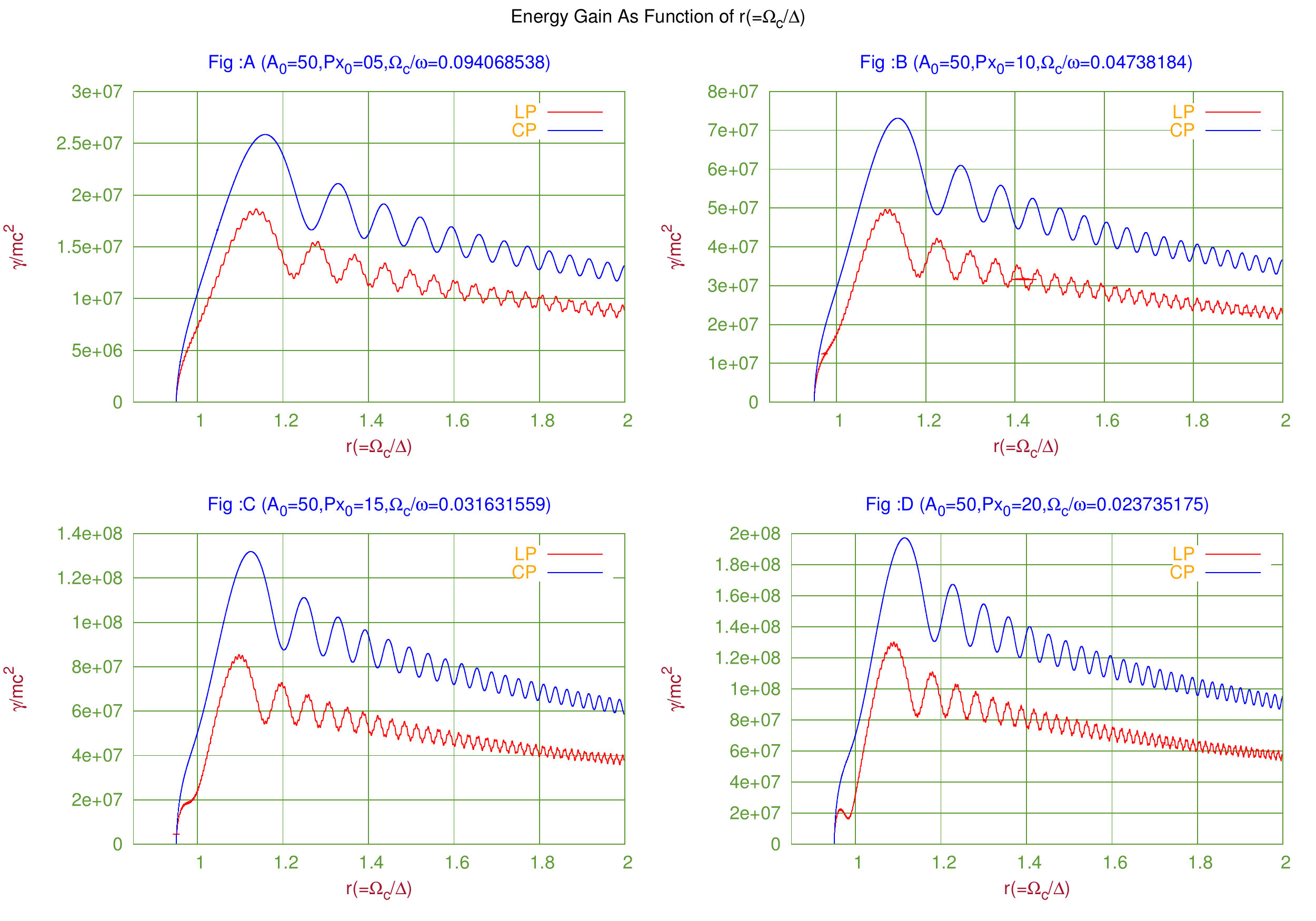}
\caption{ Energy gain of the initially non-resonant particle as function of resonant parameter $\lq r \rq$ for linearly polarizied (LP) and cricularly polarizied laser (CP).}
\end{center}
\end{figure}

\newpage
\begin{figure}
\begin{center}
\includegraphics[angle=-90, width=1\textwidth]{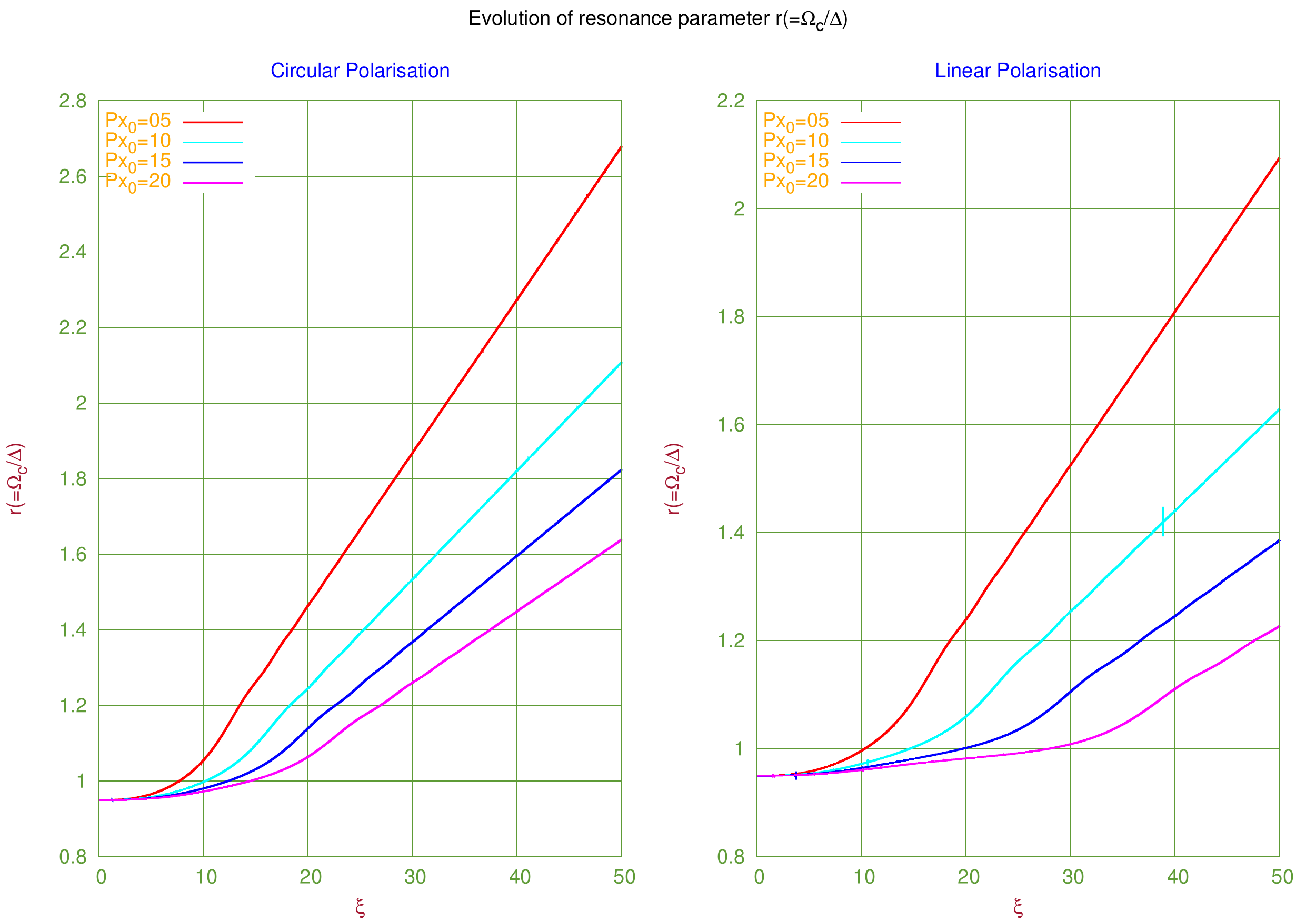}
\caption{Evolution of resonant parameter $\lq r \rq$  for different initial values of Longitudinal momentum for lineraly (LP) and circularly polarizied (CP) lasers. 
}
\end{center}
\end{figure}

\newpage
\begin{figure}
\begin{center}
\includegraphics[angle=-90, width=1\textwidth]{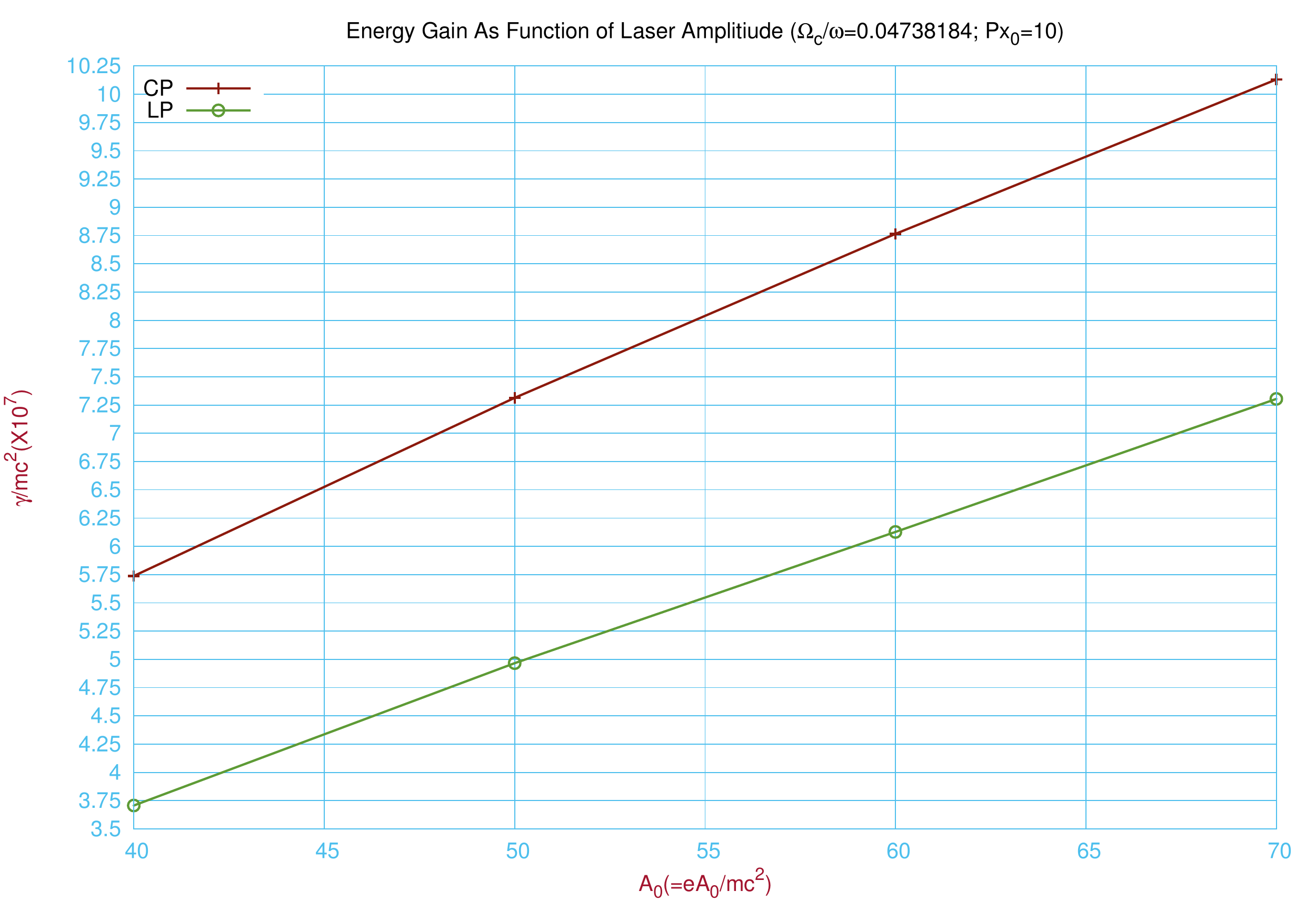}
\caption{Peak energy gain as a function of laser amplitude ($A_{0}$).}
\end{center}
\end{figure}

\newpage
\begin{figure}
\begin{center}
\includegraphics[angle=-90, width=1\textwidth]{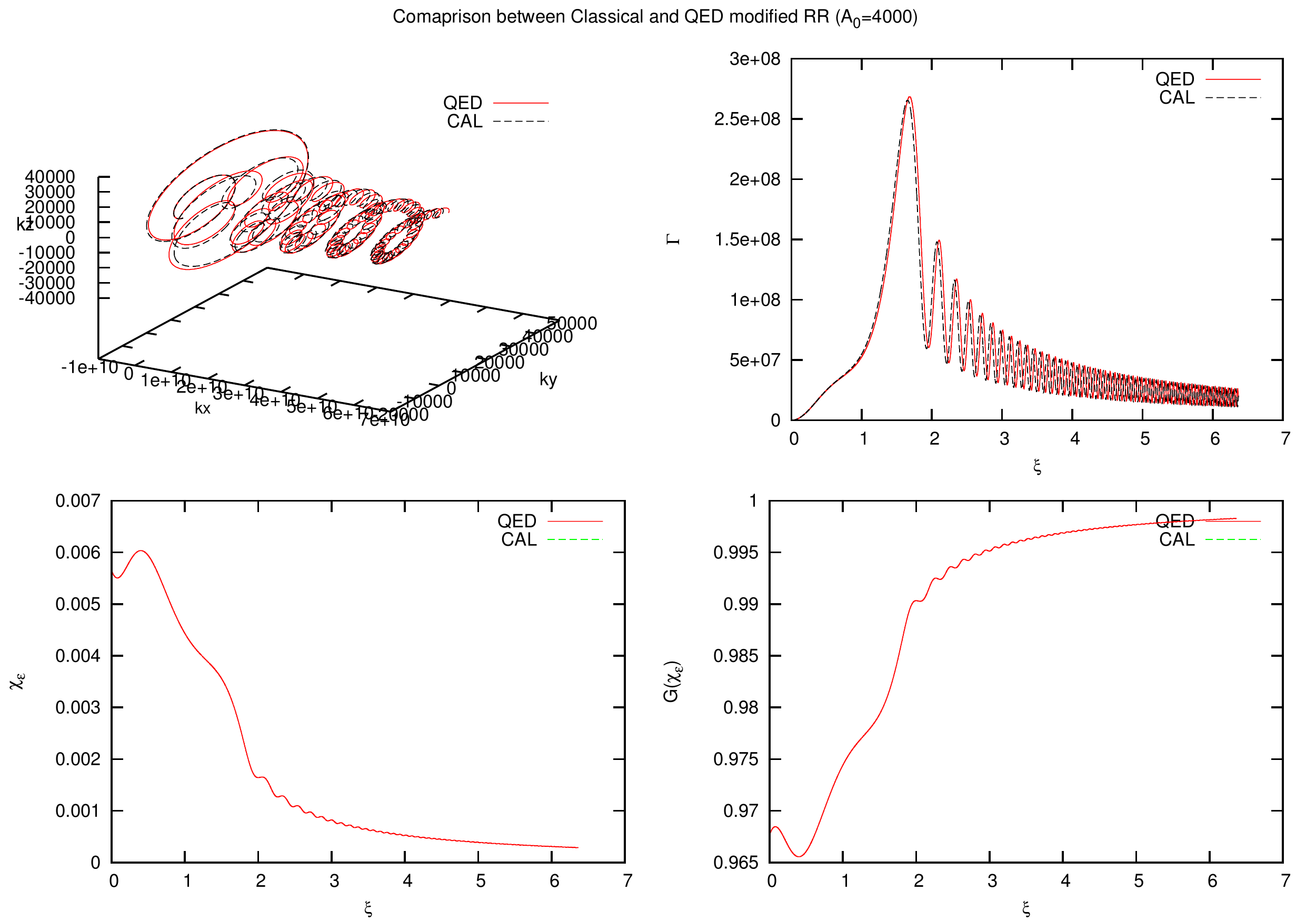}
\caption{The comparison of classical and quantum corrected energy estimates.}
\end{center}
\end{figure}

\end{document}